\documentclass[twocolumn]{jpsj3}
\usepackage{ulem}
\usepackage{braket}
\usepackage[dvipdfmx]{graphicx}
\usepackage{amsmath}
\usepackage{amssymb}
\usepackage{bm}
\usepackage{color}
\usepackage{graphicx}
\newcommand{\red}[1]{\textcolor{black}{#1}}

\definecolor{darkgreen}{rgb}{0,0.5,0}

\definecolor{purple}{rgb}{0.5,0,0.5}

\title
{Fermi Machine \\
--- Quantum Many-Body Solver Derived from Correspondence \\
between Noninteracting and Strongly Correlated Fermions } 
\author
{ Masatoshi Imada}
\inst{Physics Division, Sophia University, Chiyoda, Tokyo 102-8554, Japan \\
Faculty of Engineering, University of Tokyo, 7-3-1 Hongo, Bunkyo-ku, Tokyo 113-8656, Japan}
\abst
{Stimulated by the successful descriptions of strongly correlated electron systems by fractionalized fermions, correspondence between interacting fermions and non-interacting multi-component fermions is formulated in examples of the Hubbard model. The formalism enables  constructions of the neural network  for  quantum many-body solvers represented by coupled noninteracting fermions.  After showing the exact correspondences of 1- and 2-site Hubbard models to two-component noninteracting fermions, a numerical algorithm of the quantum machine learning for the Hubbard model is proposed. Benchmark for the 4-site systems is successfully presented and promising future directions as well as implications are discussed.  }

\begin{document}
\maketitle
\section{Introduction}
Fractionalization of electrons in strongly correlated electron systems has been proposed in several different phenomena. One dimensional itinerant electron systems exhibit the separation of spin and charge degrees of freedom~\cite{Tomonaga1950,Luttinger1963}. Models of polyacetylene accommodate spin and charge solitons with fractionalized charge as elementary excitations~\cite{SSH_RMP}. Fractional quantum Hall states under the stong magnetic fields~\cite{LaughlinFQH1983} as well as under the emergent chiral symmetry breaking~\cite{Haldane1988} show the fractionalization of electronic charge  and fractional quantizations. Slave bosons and fermions were proposed to approximately describe the 2D systems as models of cuprate high-$T_c$ superconductors~\cite{SachdevAnnPhys2003,KotliarSlaveBoson1986}. Recently, the fractionalization of an electron into two or multiple fermion components has been examined and has successfully accounted for otherwise puzzling spectroscopic experimental results of the cuprates~\cite{Sakai2016a, Imada2019,Imada2021a,Imada2021b,Yamaji2021,Singh2022}.

Meanwhile, numerical method to construct ground state wave functions of strongly interacting quantum many-body systems is one of the grand challenges in physics. Various algorithms such as a variety of auxiliary-field quantum Monte Carlo~\cite{Blankenbecler1981,SorellaQMC1989,ImadaHatsugai1989}, variational Monte Carlo~\cite{Gros1988,Shiba1988,SorellaVMC2005,Tahara-Imada2008,Misawa_VMCreview2019,Shiwei_Zhang}, density matrix renormalization group~\cite{White1992}, tensor network~\cite{Cirac2008,Orus2014,TaoXiang2012}, density matrix embedding~\cite{Garnet_ChanDMET2012} and dynamical mean-field theory~\cite{Vollhardt,Georges_RMP} have been proposed. Recently, neural network with Boltzmann machine has established an accurate way to approximate ground states of quantum spin systems~\cite{Carleo2017}  as well as itinerant fermion systems~\cite{Nomura2017}. Vision transformer, another architecture of the neural network, has also shown the state-of-the-art accuracy for quantum spin models~\cite{Viteritti2023,Rende2023}. 
We note that these neural networks are constructed by introducing hidden variables that are described by classical degrees of freedom.  

In this paper, we discuss mapping of the Hubbard model (or in more general, interacting lattice fermions) to noninteracting multi-component fermion models in the ground states that materializes the fractionalization, from which we propose a quantum machine learning algorithm aiming at an efficient quantum many-body solver.  Here, we introduce a fermion system as the hidden part coupled to the physical system, instead of the classical one such as the Ising spins in the previous neural networks. 

In Secs. I and II, the equivalence between the Hubbard model and the two-component fermion model (TCFM) is shown for the atomic limit and for the 2-site systems, respectively. Based on this equivalence, we propose a quantum machine learning algorithm in Sec. III, where the Hubbard model is approximated by a deep-layer representation of  noninteracting multi-component fermion models. The couplings between  the layers are represented by the hybridization of fermions belonging to neighboring layers. This method is viewed as an extension of restricted Boltzmann-machine representation of the neural quantum states proposed by Carleo and Troyer~\cite{Carleo2017} to a quantum neural network, where the Ising hidden variable in the Boltzmann machine is replaced by the fermionic operators that allows hybridization to generate quantum entanglement. Furthermore, it is  also regarded as an embodiment of the electron fractionalization~\cite{Imada2019,Imada2021a}, which has succeeded in solving a number of experimental puzzles in the cuprate high-$T_c$ superconductors~\cite{Sakai2016a,Yamaji2021,Imada2021a,Singh2022}. This quantum neural network can also be viewed as a tractable realization of the ground states by a systematic expansion to take account of correlation effects similarly to the formulation by the equation of motion method or equivalently the continued fraction expansion of the Green's function for the excited states~\cite{Roth,Onoda2003,Sakai2016b}. The present method does not seem to have difficulty of exponentially increasing terms encountered in the conventional equation of motion method. 

%%%%%%%%%%%%%%%%%%%%%%%%%%%%%%%%%%%%%%%%%%%%%%%%%%%%%%%%%%%%%%%%%%%%%%%%%%%%%
\section{Hubbard model in the atomic limit}
\label{1site_Hubbard}
In this section, the result shown in Appendix A of Ref.~\citen{Imada2019} is briefly summarized for the self-contained presentation.
We consider the Hubbard $U$ term 
\begin{eqnarray}
{\cal H}_{U}&=& 
Un_{\uparrow}n_{\downarrow},
\label{eq:atomic_limit}
\end{eqnarray}
with $n_{\sigma}=c^{\dagger}_{\sigma}c_{\sigma}$ for the creation (annihilation) operators $c_{\sigma}^{\dagger}$ ($c_{\sigma}$) of the spin $\sigma$,
and introduce an auxiliary fermion 
\begin{eqnarray}
\tilde{d}_{\sigma}=c_{\sigma}(1-2n_{-\sigma}),
\label{eq:d}
\end{eqnarray}
together with 
\begin{eqnarray}
\tilde{c}_{\sigma}=c_{\sigma}.
\label{eq:c}
\end{eqnarray} 
Equation (\ref{eq:atomic_limit}) can then be rewritten in a form of the TCFM as
\begin{eqnarray}
{\cal H}_{\rm TCFM}&=& {\cal H}^{(\tilde{c})} +{\cal H}^{(\tilde{d})}+{\cal H}^{(\tilde{c}\tilde{d})}, \label{eq:HTCFM} \\
{\cal H}^{(\tilde{c})}&=&\mu_{\tilde{c}} \tilde{c}_{\sigma}^{\dagger}\tilde{c}_{\sigma}, \\
{\cal H}^{(\tilde{d})}&=& \mu_{\tilde{d}}\tilde{d}_{\sigma}^{\dagger}\tilde{d}_{\sigma}, \\
{\cal H}^{(\tilde{c}\tilde{d})}&=& \Lambda (\tilde{c}_{\sigma}^{\dagger}\tilde{d}_{\sigma} + {\rm H.c}),
\end{eqnarray}
with the mapping
\begin{eqnarray}
\mu_{\tilde{c}} =\mu_{\tilde{d}}=-\Lambda=\frac{U}{2}.
\label{eq:mapping1}
\end{eqnarray}
For the derivation of Eq.(\ref{eq:HTCFM}) from Eq.~(\ref{eq:atomic_limit}), see below. 

Note first that $\tilde{d}$ and $\tilde{c}$ satisfy the exact anticommutation relation in the ground state average,
\begin{eqnarray}
\langle \tilde{c}_{\sigma}\tilde{c}_{\sigma}^{\dagger}+\tilde{c}_{\sigma}^{\dagger}\tilde{c}_{\sigma}\rangle=1, \\
\langle \tilde{d}_{\sigma}\tilde{d}_{\sigma}^{\dagger}+\tilde{d}_{\sigma}^{\dagger}\tilde{d}_{\sigma}\rangle=1, \\
\langle \tilde{c}_{\sigma}\tilde{d}_{\sigma}^{\dagger}+\tilde{d}_{\sigma}^{\dagger}\tilde{c}_{\sigma}\rangle=0,
\label{eq:anticom}
\end{eqnarray}
where
$\langle \cdots \rangle$ is defined by
\begin{eqnarray}
\langle \cdots \rangle =\frac{\langle \Phi_{0\uparrow}^{(1)} |\cdots |\Phi_{0\uparrow}^{(1)}\rangle+\langle \Phi_{0\downarrow}^{(1)} |\cdots |\Phi_{0\downarrow}^{(1)}\rangle}{\langle \Phi_{0\uparrow}^{(1)} |\Phi_{0\uparrow}^{(1)}\rangle+\langle \Phi_{0\downarrow}^{(1)} |\Phi_{0\downarrow}^{(1)}\rangle},
\label{eq:avearge}
\end{eqnarray}
because of the ground state degeneracy of Kramers doublet,
when the one-particle ground state is described  as we prove later as
\begin{equation}
|\Phi_{0\sigma}^{(1)}\rangle= ( \alpha_{c}\tilde{c}_{\sigma}^{\dagger}+\alpha_{d}\tilde{d}_{\sigma}^{\dagger} ) |0 \rangle
\label{eq:Phi01}
\end{equation}
with $\sigma$ being either $\uparrow$ or $\downarrow$.
(Note that the one-particle state must be degenerate between $|\Phi_{0\uparrow}^{(1)}\rangle$ and $|\Phi_{0\downarrow}^{(1)}\rangle$.)
In this sense, $\tilde{c}$ and $\tilde{d}$ behave as orthogonal fermions for single-particle excitations from the ground state, which is the degenerate ensemble of $|\Phi_{0\uparrow}^{(1)}\rangle$ and $|\Phi_{0\downarrow}^{(1)}\rangle$.
%Since the up spin and down spin states are degenerate and are decoupled, one can discuss only one of the spin states as spinless fermions and total wavefunction is just the product state of the two. 

By diagonalizing Eq.(\ref{eq:HTCFM}) for the case $\mu_{\tilde{c}}=\mu_{\tilde{d}}$, one obtains for the spin $\sigma$ part
\begin{eqnarray}
{\cal H}_{\rm DTCFM}&=& \mu_a a_{\sigma}^{\dagger}a_{\sigma}+\mu_b b_{\sigma}^{\dagger}b_{\sigma},  \label{eq:DHTCFM}\\
\mu_b &=& \frac{1}{2}(\mu_{\tilde{c}}+\mu_{\tilde{d}})+\Lambda, \\
\mu_a &=& \frac{1}{2}(\mu_{\tilde{c}}+\mu_{\tilde{d}})-\Lambda
\end{eqnarray}
with 
\begin{eqnarray}
b_{\sigma}&=& \frac{1}{\sqrt{2}}(\tilde{c}_{\sigma}+\tilde{d}_{\sigma})=\sqrt{2}c_{\sigma}(1-n_{-\sigma}), \label{eq:ab0}  \\
a_{\sigma}&=& \frac{1}{\sqrt{2}}(\tilde{c}_{\sigma}-\tilde{d}_{\sigma})=\sqrt{2}c_{\sigma}n_{-\sigma}.
\label{eq:ab} 
\end{eqnarray} 
The bonding and antibonding states are given by Eqs. (\ref{eq:ab0}) and (\ref{eq:ab}), respectively, which are nothing but the operators for the lower and upper Hubbard levels, respectively, in the Hubbard model whose averaged energies are given by $E=0$ and $U$. Namely, aside from the normalization factor, ${b}_{\sigma}^{\dagger}$ creates  an electron with the spin $\sigma$ when it is not occupied by the opposite spin electron (namely it fills the lower Hubbard state in the corresponding Hubbard model) as is apparent from the last equation of Eq.(\ref{eq:ab0}).
On the other hand, $a_{\sigma}^{\dagger}$, creates  an electron with the spin $\sigma$, when the opposite spin electron already exists, namely it creates a doublon at the upper Hubbard state in the Hubbard model.
It helps an intuitive interpretation of the correspondence between the Hubbard model and the TCFM.
%The lower (upper) Hubbard represents the singly (doubly) occupied particle as one sees the last expressions in Eqs.(\ref{eq:ab0}) and (\ref{eq:ab}). 
With the choice Eq.(\ref{eq:mapping1}),
$\mu_a=0$ and $\mu_b=U$ are obtained and the Hubbard gap is reproduced.
In addition, the ground state $b_{\sigma}^{\dagger}|0\rangle$ indeed satisfies the form of Eq.(\ref{eq:Phi01}).
Then the mapping between Eqs.(\ref{eq:atomic_limit}) and (\ref{eq:HTCFM})
becomes exact. In this correspondence, the Mott gap in the Hubbard model is interpreted as the hybridization gap in the TCFM.

In general, one can show that the single-particle Green's function for  the TCFM is given by
\begin{eqnarray}
 G_{\tilde{c}_{\sigma},\tilde{c}_{\sigma}^{\dagger}} (\omega ) &=& \frac{1}{\omega-\mu_{\tilde{c}}-\frac{\Lambda^2}{\omega-\mu_{\tilde{d}}}},
\nonumber \\
G_{\tilde{c}_{\sigma},\tilde{d}_{\sigma}^{\dagger}}(\omega) &=& G_{\tilde{d}_{\sigma},\tilde{c}_{\sigma}^{\dagger}}(\omega) 
=\frac{-\Lambda}{(\omega-\mu_{\tilde{c}})(\omega-\mu_{\tilde{d}})-\Lambda^2}, \nonumber \\
G_{\tilde{d}_{\sigma},\tilde{d}_{\sigma}^{\dagger}}(\omega)&=& \frac{1}{\omega-\mu_{\tilde{d}}-\frac{\Lambda^2}{\omega-\mu_{\tilde{c}}}}.
\label{eq:Gtildectilded}
\end{eqnarray}

Then from Eq.(\ref{eq:mapping1}), we obtain
\begin{eqnarray}
 G_{\tilde{c}_{\sigma},\tilde{c}_{\sigma}^{\dagger}} (\omega ) &=& \frac{1}{\omega-\frac{U}{2}-\frac{\frac{U^2}{4}}{\omega-\frac{U}{2}}} \\
 &=&\frac{1}{2}\left[ \frac{1}{\omega}+\frac{1}{\omega-U}\right].
\label{eq:G_ctilde}
\end{eqnarray}
This is equivalent to the Green's function of the atomic Hubbard, Eq.(\ref{eq:atomic_limit}).
Then the self-energy has the correct form as well:
\begin{eqnarray}
{\Sigma}(\omega)&=&\frac{\frac{U^2}{4}}{\omega-\frac{U}{2}}.
\label{eq:Sigma}
\end{eqnarray}

In this way, exact correspondence is established between the TCFM (\ref{eq:HTCFM}) and the Hubbard model (\ref{eq:atomic_limit}) in the atomic limit for the half-filled ground state as well as for single-particle excitations from it. Namely, the full Hilbert space of the Hubbard model in the atomic limit is equivalent to that of the TCFM. 

%%%%%%%%%%%%%%%%%%%%%%%%%%%%%%%%%%%%%%%%%%%%%%%%%%%%%%%%%%%%%%%%%%%%%%%%%%%%%%%%

\section{Two-site Hubbard model}
\label{2site_Hubbard}

In this section, the mapping of the atomic limit already established and summarized in the last section is extended and a mapping between the 2-site Hubbard model and the 2-site TCFM is shown.

The 2-site Hubbard Hamiltonian reads 
\begin{eqnarray}
{\cal H}&=&{\cal H}_{t} + {\cal H}_{U}, \label{eq:2siteHubbard}\\
{\cal H}_{t} &=& \sum_{\sigma}[ -t(c_{1\sigma}^{\dagger}c_{2\sigma}+c_{2\sigma}^{\dagger}c_{1\sigma})+\mu\sum_{i=1,2}n_{i,\sigma}], 
\label{eq:2siteHubbardt}\\
{\cal H}_{U}&=& 
%U(n_{\uparrow}-\frac{1}{2})(n_{\downarrow}-\frac{1}{2}),
U\sum_{i=1,2}n_{i\uparrow}n_{i\downarrow}.
\label{eq:2siteHubbardU}
\end{eqnarray}
with $n_{i,\sigma}=c^{\dagger}_{i,\sigma}c_{i,\sigma}$. We take $\mu=0$, because in the canonical ensemble, spatially uniform chemical potential does not change physical properties except for a trivial energy shift.

We first analyze the half-filled case with one $\uparrow$ and one $\downarrow$ electron and use the basis of the full Hilbert space expanded by $|\uparrow,\downarrow\rangle, |\downarrow,\uparrow\rangle, |\uparrow\downarrow,0\rangle, |0, \uparrow\downarrow\rangle$ in the notation $|n_{1\uparrow}n_{1\downarrow}, n_{2\uparrow}n_{2\downarrow}\rangle$,
where $n_{i\sigma}$ is the number of spin $\sigma$ fermion at the site $i$ and is denoted as $\uparrow$ if $n_{i\uparrow}=1$ and $\downarrow$ if $n_{i\downarrow}=1$ while $0$ if  $n_{i\uparrow}=n_{i\downarrow}=0$ at the site $i$. For instance, $|\uparrow,\downarrow\rangle$ represents 
$c_{1\uparrow}^{\dagger}c_{2\downarrow}^{\dagger}|0\rangle$ and $|\downarrow,\uparrow\rangle=c_{2\uparrow}^{\dagger}c_{1\downarrow}^{\dagger}|0\rangle$ such that  up-spin creation operators are ordered in the left-hand side and spatial sites are ordered in the site-number order from 1 to 2 for the same spin. Here $|0\rangle$ is the vacuum. In this notation, the Hamiltonian matrix is written as
\begin{eqnarray}
{\cal H}=\left(
\begin{array}{c|cccc} 
& |\uparrow,\downarrow\rangle & |\downarrow,\uparrow\rangle & |\uparrow\downarrow,0\rangle & |0, \uparrow\downarrow\rangle \\ \hline
\langle\uparrow,\downarrow|  & 0 & 0 & -t & -t \\
\langle\downarrow,\uparrow| & 0 & 0 & -t & -t \\
\langle \uparrow\downarrow, 0| &-t &-t & U & 0\\
 \langle 0, \uparrow\downarrow| &-t &-t & 0 & U 
\end{array} 
\right).
\label{eq:HubbardMatrix}
\end{eqnarray}
\red{Here and hereafter, we have explicitly written the choice of the basis as the row and column vectors to make the definition of the components of the $4\times 4$ matrix clearer.}

The eigenvalues are
\begin{eqnarray}
E_0&=&\frac{U}{2}Q, \nonumber\\ 
E_1&=& 0, \nonumber\\
E_2&=& U, \nonumber\\ 
E_3&=& \frac{U}{2}P, 
\label{eq:Eigenvalue}
\end{eqnarray}
where $P=1+\sqrt{1+R^2}$, $Q=1-\sqrt{1+R^2}$ and $R=4t/U$.
The normalized eigenfunction of the ground state with the energy $E_0$ is given by
\begin{equation}
(R,R,-Q,-Q)/\sqrt{2(R^2+Q^2)},
\label{eq:eigenvector}
\end{equation}
which is the coefficients in the basis of $(|\uparrow,\downarrow\rangle, |\downarrow,\uparrow\rangle, |\uparrow\downarrow,0\rangle, |0,\uparrow\downarrow\rangle)$.
Namely, the normalized ground state is
\begin{equation}
\frac{1}{\sqrt{2(R^2+Q^2)}}(R|\uparrow,\downarrow\rangle+R|\downarrow,\uparrow\rangle- Q|\uparrow\downarrow,0\rangle- Q|0,\uparrow\downarrow\rangle ).
\label{eq:eigenfunction}
\end{equation}

Now we introduce the same auxiliary fermions Eqs.(\ref{eq:d}) and (\ref{eq:c}) as the atomic limit.
By using $\tilde{c}$ and $\tilde{d}$, the Hubbard Hamiltonian (\ref{eq:2siteHubbard}) can be mapped to a noninteracting 2-site TCFM as
\begin{eqnarray}
{\cal H}_{\rm TCFM}&=& {\cal H}^{(\tilde{c})} +{\cal H}^{(\tilde{d})}+{\cal H}^{(\tilde{c}\tilde{d})}, \label{eq:hamiltonian_TCFM} \\
{\cal H}^{(\tilde{c})}&=&\sum_{\sigma}[-t_{\tilde{c}}(\tilde{c}_{1,\sigma}^{\dagger}\tilde{c}_{2,\sigma}+{\rm H.c.} )
+\mu_{\tilde{c}}\sum_{i=1,2}n_{\tilde{c},i,\sigma}] , \ \ \ \ \\
{\cal H}^{(\tilde{d})}&=& \sum_{\sigma}[-t_{\tilde{d}}(\tilde{d}_{1,\sigma}^{\dagger}\tilde{d}_{2,\sigma}+{\rm H.c.} )
+\mu_{\tilde{d}}\sum_{i=1,2}n_{\tilde{d},i,\sigma}] ,  \ \ \ \ \\
{\cal H}^{(\tilde{c}\tilde{d})}&=& \Lambda \sum_{\sigma}\sum_{i=1,2}(\tilde{c}_{i,\sigma}^{\dagger}\tilde{d}_{i,\sigma} + {\rm H.c}),
\end{eqnarray}
where $n_{\tilde{c},i,\sigma} =\tilde{c}_{i,\sigma}^{\dagger}\tilde{c}_{i,\sigma}$ and  $n_{\tilde{d},i,\sigma} =\tilde{d}_{i,\sigma}^{\dagger}\tilde{d}_{i,\sigma}$ 
again with the same mapping Eq.~(\ref{eq:mapping1}) together with
\begin{eqnarray}
\tilde{t}_{\tilde{c}}=t, 
\tilde{t}_{\tilde{d}}=-t.
\label{eq:mapping2}
\end{eqnarray}

The Hilbert space of this TCFM wave function may be expanded in the basis of
$(\tilde{c}_{1,\sigma}^{\dagger}+\tilde{d}_{1,\sigma}^{\dagger})/\sqrt{2}, (\tilde{c}_{2,\sigma}^{\dagger}+\tilde{d}_{2,\sigma}^{\dagger})/\sqrt{2}, (\tilde{c}_{1,\sigma}^{\dagger}-\tilde{d}_{1,\sigma}^{\dagger})/\sqrt{2}, (\tilde{c}_{2,\sigma}^{\dagger}-\tilde{d}_{2,\sigma}^{\dagger})/\sqrt{2}))|0\rangle,
$
where $|0\rangle$ is the vacuum of $\tilde{c}$ and $\tilde{d}$. 
Since the spin degeneracy exists and the correlation between opposite spin fermions does not exist, we use an abbreviated notation of this basis as
$\tilde{b}_{i}=(\tilde{c}_{i}+\tilde{d}_{i})/\sqrt{2}, \tilde{a}_{i}= (\tilde{c}_{i}-\tilde{d}_{i} )/\sqrt{2}$ for $i=1,2$. The Hamiltonian matrix in this representation is
\begin{eqnarray}
{\cal H}_{\rm TCFM}=\left(
\begin{array}{c|cccc} 
&\tilde{b}_{1} & \tilde{b}_{2} & \tilde{a}_{1}& \tilde{a}_{2} \\ 
\hline
\tilde{b}_{1}^{\dagger} & 0 & 0 & -t & -t \\
\tilde{b}_{2}^{\dagger} & 0 & 0 & -t & -t \\ 
\tilde{a}_{1}^{\dagger} &-t &-t & U & 0\\
\tilde{a}_{2}^{\dagger} &-t &-t & 0 & U
\end{array} 
\right),
\label{eq:TCFMMatrix}
\end{eqnarray}
for both of up and down spin sectors, 
which is the same as Eq.(\ref{eq:HubbardMatrix}). Then the eigenvalues and eigenfunctions are of course the same as Eqs.(\ref{eq:Eigenvalue}) and (\ref{eq:eigenfunction}), respectively.
The ground-state wave function is given by 
\begin{eqnarray}
|\Phi_0^{\rm TCFM}\rangle &=&\frac{1}{2(R^2+Q^2)}(R(\tilde{b}_{1,\uparrow}^{\dagger}+\tilde{b}_{2,\uparrow}^{\dagger}) %\nonumber \\
+ Q(\tilde{a}_{1,\uparrow}^{\dagger}+\tilde{a}_{2,\uparrow}^{\dagger})) \nonumber \\
&\times& (R(\tilde{b}_{1,\downarrow}^{\dagger}+\tilde{b}_{2,\downarrow}^{\dagger}) %\nonumber \\
+ Q(\tilde{a}_{1,\downarrow}^{\dagger}+\tilde{a}_{2,\downarrow}^{\dagger})) |0\rangle,
\label{eq:TCFM_E0_wf}
\end{eqnarray}
which has the symmetries of spin singlet and spatial parity even.
The first excited state is given by
\begin{eqnarray}
|\Phi_1^{\rm TCFM}\rangle&=&\frac{\sqrt{2}}{4(R^2+Q^2)}  \nonumber \\
&\times & 
((R(\tilde{b}_{1,\uparrow}^{\dagger}-\tilde{b}_{2,\uparrow}^{\dagger}) %\nonumber \\
+ Q(\tilde{a}_{1,\uparrow}^{\dagger}-\tilde{a}_{2,\uparrow}^{\dagger}))\nonumber \\
&&\times  (R(\tilde{b}_{1,\downarrow}^{\dagger}+\tilde{b}_{2,\downarrow}^{\dagger}) %\nonumber \\
+ Q(\tilde{a}_{1,\uparrow}^{\dagger}+\tilde{a}_{2,\uparrow}^{\dagger})) \nonumber \\
 &-& (R(\tilde{b}_{1,\uparrow}^{\dagger}+\tilde{b}_{2,\uparrow}^{\dagger})
+ Q(\tilde{a}_{1,\uparrow}^{\dagger}+\tilde{a}_{2,\uparrow}^{\dagger})) \nonumber \\
 &&\times  ( R(\tilde{b}_{1,\downarrow}^{\dagger}-\tilde{b}_{2,\downarrow}^{\dagger}) %\nonumber \\
+Q( \tilde{a}_{1,\downarrow}^{\dagger}-\tilde{a}_{2,\downarrow}^{\dagger} )))|0\rangle,\nonumber \\
\label{eq:TCFM_E1_wf}
\end{eqnarray}
which is spin triplet and spatial parity odd.
Other two higher excited states can be given similarly.

With this correspondence, the TCFM ground state (Eq.(\ref{eq:TCFM_E0_wf})) is exactly mapped to the Hubbard terminology as
\begin{eqnarray}
|\Phi_0^{\rm Hub}\rangle 
&=& \frac{1}{2(R^2+Q^2)}
(R(\tilde{b}_{1,\uparrow}^{\dagger}\tilde{b}_{2,\downarrow}^{\dagger}
+\tilde{b}_{2,\uparrow}^{\dagger}\tilde{b}_{1,\downarrow}^{\dagger}) \nonumber \\
&&+Q(\tilde{a}_{1,\uparrow}^{\dagger}\tilde{b}_{1,\downarrow}^{\dagger}+\tilde{a}_{2,\uparrow}^{\dagger}\tilde{b}_{2,\downarrow}^{\dagger})) |0\rangle
\nonumber \\
&\Leftrightarrow&  
{\rm Eq.(29)}.
\label{eq:Hub_E0_wf}
\end{eqnarray}
Here, in the Hubbard model, we used the relations $(\tilde{c}_{i,\sigma}^{\dagger}-\tilde{d}_{i,\sigma}^{\dagger})|0\rangle=0$,
$(\tilde{c}_{i,\sigma}^{\dagger}+\tilde{d}_{i,\sigma}^{\dagger})(\tilde{c}_{i,-\sigma}^{\dagger}+\tilde{d}_{i,-\sigma}^{\dagger})|0\rangle=0$
and 
$(\tilde{c}_{1i,\sigma}^{\dagger}-\tilde{d}_{i,\sigma}^{\dagger})(\tilde{c}_{j,-\sigma}^{\dagger}+\tilde{d}_{j,-\sigma}^{\dagger})|0\rangle=0$
for $i\ne j$,
because $\tilde{c}_{i,\sigma}^{\dagger}-\tilde{d}_{i,\sigma}^{\dagger}$ creates a doublon at the $i$th site singly occupied by a spin $-\sigma$ electron and $\tilde{c}_{i,\sigma}^{\dagger}+\tilde{d}_{i,\sigma}^{\dagger}$ creates an electron at the empty $i$th site.
Mapping for the excited states is similarly shown.
Therefore, the exact mapping of not only the ground state but also the whole structure of the spectra between the TCFM and Hubbard model is proven for the two-site system as well.

Here, we remark on the spin degeneracy in the TCFM. Because of the spin degeneracy, the total ground state of the TCFM with one up and one down spin sector is given by the product state $|\Phi_0\rangle=|\Phi_{0\uparrow}\rangle|\Phi_{0\downarrow}\rangle$.
Then the total ground-state energy with one up and one down spin fermions is twice of the ground state energy $E_0$ in Eq.(\ref{eq:Eigenvalue}).
However, as in the single-site problem, the ground state average in the TCFM  
$\langle \cdots \rangle$ should be taken as  Eq.(\ref{eq:avearge}) and the factor 2 is canceled.

For the filling doped with one hole or one electron from the half filling has also trivially the same mapping.
For instance, the energy of  a hole doped ground state is 
\begin{eqnarray}
|\Phi_0^{\rm TCFM}\rangle &=&\frac{1}{2}(\tilde{b}_{1,\uparrow}^{\dagger}+\tilde{b}_{2,\uparrow}^{\dagger})|0\rangle,
\label{eq:TCFM_doped_E0_wf}
\end{eqnarray}
which has the energy $E_0=0$ and is the same as the one-hole doped Hubbard model.

\section{Fermi Machine as a Quantum Neural Network}
\label{sec:QML}
We have shown that the Hubbard model has an exact mapping to the TCFM for one- and two-site problems and the ground-state wave function of the  Hubbard model can be constructed from the ground state of the TCFM. It demonstrates that the TCFM is able to describe the energy scales of the Mott gap separating the lower and upper Hubbard bands  scaled by $E_2-E_0\propto U$ and the singlet-triplet gap ($E_1-E_0$ and $E_3-E_2$) associated with the superexchange interaction $J$, which scales to $4t^2/U$ in the strong coupling limit. Up to the 2 sites, the exact ground and excited states of the Hubbard model are represented by the corresponding ground and excited states of the TCFM, respectively by optimizing  the variational parameters given by Eqs.(\ref{eq:mapping1}) and (\ref{eq:mapping2}), namely, $\mu_{\tilde{c}}, \mu_{\tilde{d}}, \Lambda, t_{\tilde{c}}$ and $t_{\tilde{d}}$.

This suggests a potential to represent  $N_s$-site Hubbard models defined by 
\begin{eqnarray}
{\cal H}&=&{\cal H}_{t} + {\cal H}_{U}, \label{eq:Ns_siteHubbard}\\
{\cal H}_{t} &=& \sum_{\langle i,j \rangle,\sigma}[ -t(c_{i\sigma}^{\dagger}c_{j\sigma}+{\rm H.c.} %_{\sigma}^{\dagger}c_{1\sigma})
+\mu\sum_{i}^{N_s}n_{i,\sigma}], 
\label{eq:Ns_siteHubbardt}\\
{\cal H}_{U}&=& 
U\sum_{i}^{N_s}n_{i\uparrow}n_{i\downarrow}
\label{eq:Ns_siteHubbardU}
\end{eqnarray}
for any system size $N_s$ by a quantum neural network along this line. 
We will discuss the representability of the wave function later. 

In principle, by introducing the self-energies of the visible electron $c$ through the effect of the hidden fermions $d$, it was shown that the spectral function and energy spectra of any interacting system can be represented formally but exactly by the hierarchy of the self-energy structure, which 
represents the energy eigenvalues of the Hamiltonian by poles and zeros of Green's function through the continued fraction expansion~\cite{Roth,Onoda2003,Sakai2016b} as
\begin{eqnarray}
G(q,\omega)&=&\frac{1}{\omega-\epsilon_c(q)-\Sigma_1(q,\omega)}, \label{eq:continued_fraction_Green's_func}\\
\Sigma_1(q,\omega)&=&\frac{\eta_1(q)}{\omega-\epsilon_1(q)-\Sigma_2(q,\omega)}, \nonumber \\
\Sigma_2(q,\omega)&=&\frac{\eta_2(q)}{\omega-\epsilon_2(q)-\Sigma_3(q,\omega)}. \nonumber \\
\cdots && \nonumber
\end{eqnarray}
Here we assume the translational symmetry of the Hamiltonian parameters to allow the momentum representation.
This can be achieved by formally tracing the equation of motion for the $c$ operator in the Heisenberg representation as
\begin{eqnarray}
i\hbar \frac{dc(t)}{dt}&=& [c,{\cal H}], \label{eqmotion}
\end{eqnarray}
for any Hamiltonian $\cal H$ and $[A,B]\equiv AB-BA$.

\begin{figure}[tb]
		\begin{center}
			\includegraphics[width=7cm]{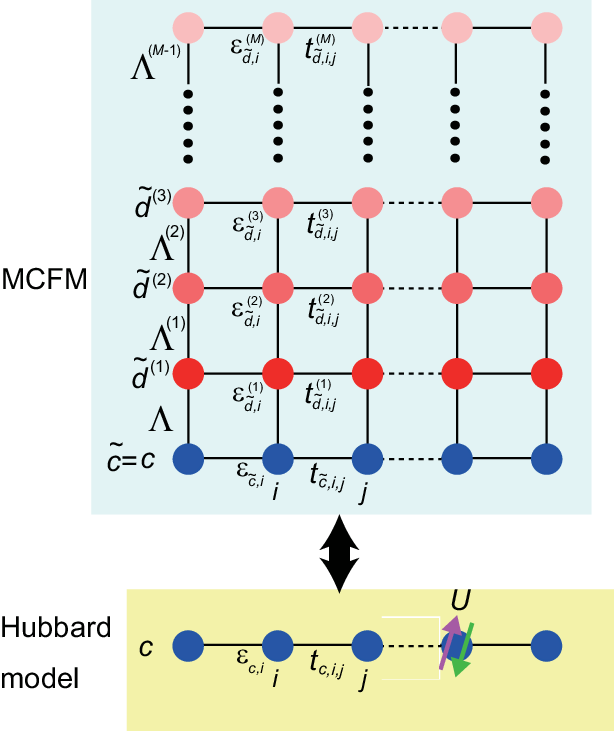}   
	\caption{(Color online): Illustration of the architecture of Fermi machine network based on the MCFM compared to the Hubbard model.}
	\label{fig:mapping_MCFM}
		\end{center}
	\end{figure}

Therefore, one can expect that the ground state of the Hubbard model (or any interacting lattice fermions) can be mapped to an optimized non-interacting multi-component fermion model (MCFM) represented by 
\begin{eqnarray}
{\cal H}_{\rm MCFM}&=& {\cal H}^{(\tilde{c})} +{\cal H}^{(\tilde{c}\tilde{d})} +{\cal H}^{(\tilde{d})}, \label{eq:MCFM_hamiltonian} \\
{\cal H}^{(\tilde{c})}&=&\sum_{\langle i,j \rangle, \sigma}(-t_{\tilde{c},i,j})(\tilde{c}_{i,\sigma}^{\dagger}\tilde{c}_{j,\sigma}+{\rm H.c.}) +\sum_{i,\sigma}\mu_{\tilde{c},i}n_{\tilde{c},i,\sigma}, \nonumber \\
\label{eq:MCFM_Hc} \\
{\cal H}^{(\tilde{c}\tilde{d})}&=& \Lambda \sum_{\sigma}\sum_{i}(\tilde{c}_{i,\sigma}^{\dagger}\tilde{d}_{i,\sigma}^{(1)} + {\rm H.c}), \label{MCFM_Hcd} \\
{\cal H}^{(\tilde{d})}&=&\sum_{m=1}^{M} [\sum_{\langle i,j \rangle \sigma}(-t_{\tilde{d},i,j}^{(m)})(\tilde{d}_{i,\sigma}^{(m) \dagger}\tilde{d}_{j,\sigma}^{(m)}+{\rm H.c.}) \nonumber \\
&&+\sum_{i,\sigma}\mu_{\tilde{d},i}^{(m)}\tilde{d}_{i,\sigma}^{(m) \dagger}\tilde{d}_{i,\sigma}^{(m)}]  \nonumber \\
&+&\sum_{m=1}^{M-1}\Lambda^{(m)}\sum_{i,\sigma}(\tilde{d}_{i,\sigma}^{(m) \dagger}\tilde{d}_{i,\sigma}^{(m+1)}+{\rm H.c.}), \label{eq:MCFM_Hd} 
\end{eqnarray}
as is illustrated in Fig.~\ref{fig:mapping_MCFM},
because the Green's function of Eq.~(\ref{eq:MCFM_hamiltonian}) is given by the hierarchical structure given by Eq.(\ref{eq:continued_fraction_Green's_func}) with the correspondence 
\begin{eqnarray}
\epsilon_c(q)&=&\frac{1}{N_s}\sum_{\langle \ell,j\rangle}e^{iq(\ell-j)} (-t_{\tilde{c},\ell,j})+\mu_{\tilde{c}},%+\sum_{\ell}e^{-iq\ell}\mu_{\tilde{c},\ell}, 
\nonumber \\
\eta_1(q)&=&\Lambda, \nonumber \\
\epsilon_{m}(q)&=& \frac{1}{N_s} \sum_{\langle \ell,j \rangle}e^{iq\ell-j)}(-t_{\tilde{d},\ell,j}^{(m)}) + \mu_{\tilde{d},i}^{(m)}, \nonumber \\
\eta_{m}(q)&=&\Lambda^{(m)}, \label{eq:MCFM_Hubbard} 
\end{eqnarray}
if the chemical potential is spatially uniform.

The algorithm of the present quantum neural network to represent the ground state of the Hubbard model is the following:
\begin{itemize}
\item
We diagonalize the noninteracting MCFM Hamiltonian Eq.~(\ref{eq:MCFM_hamiltonian}) and obtain the ground state by filling the low-energy fermions up to the Fermi level for a given set of parameters in Eqs.~(\ref{eq:MCFM_Hc})-(\ref{eq:MCFM_Hd}). 
\item
Next we replace the hidden fermions $\tilde{d}$ by using the rule %Eqs.~(\ref{eq:mapping1}), (\ref{eq:mapping2}) and 
Eq.(\ref{eq:d}) together with similar correspondence such as $\tilde{d}_{i,\sigma}^{(m)}=\tilde{d}_{i,\sigma}^{(m-1)}(1-2n_{\tilde{d}^{(m-1)},i,-\sigma})$
where $n_{\tilde{d}^{(m)},i,-\sigma} =\tilde{d}_{i,-\sigma}^{(m)\dagger}\tilde{d}_{i,-\sigma}^{(m)}$.
\item
Then by using the relations such as those below Eq.~(\ref{eq:Hub_E0_wf}) used to derive Eq.~(\ref{eq:Hub_E0_wf}) and its extension to deep-layer hidden operators, one can represent the wave function for the Hubbard model $|\Phi_0^{\rm Hub}\rangle$ as in the derivation of Eq.~(\ref{eq:Hub_E0_wf}), from which one can easily calculate the matrix elements of the Hubbard model and the energy expectation value of the Hubbard model by 
 \begin{eqnarray}
\langle E_0\rangle =\langle {\cal H} \rangle =\frac{\langle \Phi_0^{\rm Hub}|{\cal H}|\Phi_0^{\rm Hub}\rangle}{\langle \Phi_0^{\rm Hub}|\Phi_0^{\rm Hub} \rangle }
\label{eq:VarEnergy}
\end{eqnarray}
using Monte Carlo sampling of the real space configuration of $c$ fermion occupation.
Here $\cal H$ is given by Eqs.~(\ref{eq:Ns_siteHubbard})-(\ref{eq:Ns_siteHubbardU}).
Note that ${\cal H}_t|\Phi_0^{\rm Hub}\rangle$ is not a single Slater determinant even when $|\Phi_0^{\rm Hub}\rangle$ is so. An easy way to obtain the average is 
\begin{eqnarray}
\langle E_{0t}\rangle =\langle {\cal H}_t \rangle=\lim_{\tau \rightarrow 0} \frac{1}{\tau}\log \left[\frac{\langle \Phi_0^{\rm Hub}|\exp[\tau{\cal H}_t]|\Phi_0^{\rm Hub}\rangle}{\langle \Phi_0^{\rm Hub}|\Phi_0^{\rm Hub} \rangle }\right]
\label{eq:VarEnergy2}
\end{eqnarray}
because $\exp[\tau{\cal H}_t]|\Psi\rangle$ remains a single Slater determinant when $|\Psi\rangle$ is a single Slater determinant~\cite{ImadaHatsugai1989}.
\item
Finally, by using the variational principle, the parameters in the MCFM Hamiltonian Eqs.~(\ref{eq:MCFM_hamiltonian})-(\ref{eq:MCFM_Hd})  are regarded as the variational parameters and are optimized to lower the energy Eq.~(\ref{eq:VarEnergy}) by following the variational principle.
\end{itemize}

There exist a few useful relations to simplify the MCFM state:
\begin{eqnarray}
\tilde{d}_{i,\sigma}^{(m)}&=&\tilde{d}_{i,\sigma}^{(m-1)}(1-2n_{\tilde{d}^{(m-1)},i,-\sigma})=\cdots \nonumber \\
&=&c_{i,\sigma}(1-2n_{c_{i,-\sigma}})^{m},  \label{eq:d-correspondence1} \\
n_{\tilde{d}_{i,\sigma}^{(m)}}&=&n_{\tilde{d}^{(m-1)}_{i,\sigma}}=\cdots=n_{c_{i,\sigma}} 
\label{eq:d-correspondence2}
\end{eqnarray}
The ground state of the MCFM can be represented generally as 
\begin{eqnarray}
|\Psi_0^{\rm MCFM}\rangle&=& 
\begin{pmatrix} |\Psi_{0 \ \uparrow}^{\rm MCFM}\rangle & 0 \\ 0 & |\Psi_{0 \ \downarrow}^{\rm MCFM}\rangle 
\end{pmatrix},  \label{eq:MCFM_ground_state0} \\
|\Psi_{0 \ \sigma}^{\rm MCFM}\rangle&=& \prod_q^{k_{\rm F}} |\Psi_{\sigma}^{\rm MCFM}(q)\rangle \label{eq:MCFM_ground_state}, \\
|\Psi_{\sigma}^{\rm MCFM}(q)\rangle&=& [\alpha_{0,q}\tilde{c}_{q,\sigma}^{\dagger}+\sum_{m=1}^{M}\alpha_{m,q}\tilde{d}_{m,q,\sigma}^{\dagger}]|0\rangle, 
\label{eq:Psi_MCFM_qsigma}
\end{eqnarray}
where $q$ is the momentum.
Here, in the ground state of the MCFM,  
fermions 
occupy up to the fermi momentum $k_{\rm F}$. 
Since the MCFM is diagonal in the momentum space, the ground-state wave function in the $k$ space is trivial if the parameters in Eqs.~(\ref{eq:MCFM_Hc})-(\ref{eq:MCFM_Hd}) are given. The coefficient $\alpha_{m,q}$ ($m=0,\cdots M$) is determined from the eigenvector of the ground state, which can be easily calculated numerically.

This MCFM ground state is mapped to the variational form of the ground state of the Hubbard model as
\begin{eqnarray}
|\Psi_0^{\rm Hub}\rangle&=& \begin{pmatrix} |\Psi_{0 \ \uparrow}^{\rm Hub}\rangle & 0 \\ 0 & |\Psi_{0 \ \downarrow}^{\rm Hub}\rangle \end{pmatrix}, \label{eq:Psi_matrix}
\end{eqnarray}
\begin{equation}
|\Psi_{0 \ \sigma}^{\rm Hub}\rangle=\prod_q^{k_{\rm F}}[\sum_{m=0}^{M}\alpha_{m,q}\sum_{\ell}\frac{e^{iq\ell}}{\sqrt{N_s}}c_{\ell,\sigma}^{\dagger}(1-2n_{\ell,-\sigma})^m]|0\rangle.
\label{eq:Hub_var_ground_state}
\end{equation}
It can be rewritten as
\begin{eqnarray}
|\Psi_{0 \ \sigma}^{\rm Hub}\rangle &=& 
\prod_q^{k_{\rm F}}|\Psi_{\sigma}^{\rm Hub}(q)\rangle,
\label{eq:Hub_var_ground_state2} 
\end{eqnarray}
\begin{equation}
|\Psi_{\sigma}^{\rm Hub}(q)\rangle = 
[\sum_{m=0}^{M}\alpha_{m,q}\sum_{\ell}\frac{e^{iq\ell}}{\sqrt{N_s}}c_{\ell,\sigma}^{\dagger}(-)^{mn_{\ell,-\sigma}}]|0\rangle. \label{eq:Psi_Hub_qsigma}
%\nonumber 
\end{equation}
It should be noted that $|\Psi_{\sigma}^{\rm Hub}(q)\rangle$ contains electron operators $c^{\dagger}_{p,\sigma}$ with not only the momentum $q$ but also all the momentum $p$ in the Brillouin zone in the momentum representation, which is a consequence of the interference with the spin $-\sigma$ electron at the $\ell$-th site in Eq.(\ref{eq:Psi_Hub_qsigma}) reflecting the scattering of $\sigma$ and $-\sigma$ electrons at the $\ell$-th site in the original Hubbard model.  Indeed, intuitively, the minus factor for the odd $m$ in Eq.~(\ref{eq:Psi_Hub_qsigma}) represents the reduction of the weight of $c_{\ell,\sigma}^{\dagger}$ electron if $-\sigma$ electron exists at the $\ell$-th site to punish the double occupation
leading to the entanglement of spin up and down electrons.
\red{For instance, in the large $U$ limit, the corresponding $\Lambda \rightarrow -\infty$ makes the perfect exclusion of the double occupation through Eq.~(\ref{eq:Psi_Hub_qsigma}).}
The representability of this wave function is discussed in Appendix. 

To numerically optimize the variational parameters in the MCFM (namely the parameters in Eqs.~(\ref{eq:MCFM_hamiltonian})-(\ref{eq:MCFM_Hd})), we need to estimate the ground-state energy using the obtained wave function through  Eq.~(\ref{eq:VarEnergy}),
by the average of the inserted sample $|x\rangle$ as
\begin{eqnarray}
E_0&=&\langle {\cal H} \rangle=\sum_x\rho(x)F(x), \label{eq:E_0_MC} \\
\rho(x)&=&|\langle x|\Psi_0^{\rm Hub} \rangle|^2/\langle\Psi_0^{\rm Hub} |\Psi_0^{\rm Hub} \rangle, \label{eq:E_0_MC} \\
F(x)&=& \sum_{x'}\frac{\langle \Psi_{0}^{\rm Hub}|x'\rangle}{\langle \Psi_{0}^{\rm Hub}|x\rangle}\langle x'| {\cal H}|x\rangle.
\label{eq:E_0ave} 
\end{eqnarray}
Here, $\langle x'| {\cal H}_t|x\rangle$ can be calculated in the same way as Eq.~(\ref{eq:VarEnergy2}).

Hereafter, let us consider a $N_s$-site system with $N/2$ up spin and down spin electrons each. (Extension to the case of an arbitrary number of up and down spin electrons, $N_{\uparrow}$, and $N_{\downarrow}$, respectively, is easy and straightforward.) 
The Monte Carlo sampling can be performed by taking $\rho(x)$ as the probability to generate samples by following  the importance sampling and the simple average of $F(x)$ over sampling gives us the estimate of $E_0$.   
Here, $|x\rangle$ is a sample of real space configuration (simple product state), such as $(0,\uparrow,\uparrow,\uparrow \downarrow, \downarrow, 0, \cdots)$, namely
\begin{eqnarray}
|x\rangle&=&\prod_{\ell}^{N/2}c_{\ell,\uparrow}^{\dagger}\prod_{\ell'}^{N/2}c_{\ell',\downarrow}^{\dagger}|0\rangle,
\label{eq:x-sample}
\end{eqnarray}
where the sets of the real space coordinate $\{ \ell \}$ and $\{ \ell' \}$ specify the positions of the up and down spin electrons, respectively in the given sample. 
Since $\cal H$ is a sparse matrix by assuming the system with the short-ranged hoppings and interactions, and has nonzero matrix element only in their ranges, the summation over $|x'\rangle$ can be explicitly taken for each $|x\rangle$.

To enhance the representability of the wave function, one can optionally introduce the ``Fermi distribution factor" $f$ 
on top of Eq.(\ref{eq:Hub_var_ground_state}) (or Eqs.(\ref{eq:Hub_var_ground_state2}) and (\ref{eq:Psi_Hub_qsigma})) similarly to the finite-temperature Boltzmann factor in the classical Boltzmann machine.
Namely, in this scheme, $|\Psi_{0 \ \sigma}^{\rm Hub} \rangle$ is modified and replaced by
\begin{eqnarray}
|\Psi_{0 \ \sigma}^{\rm Hub}\rangle&=&
\prod_q  f(\beta,q) |\Psi_{\sigma}^{\rm Hub}(q)\rangle,
\label{eq:MCFM_var_ground_state_Fermi} \\
f(\beta,q)&=&\frac{1}{1+e^{\beta {\cal E}(q)}}, \label{eq:f_xPsi} \\
{\cal E}(q)&=&
\sum_{\sigma}
\frac{\langle \Psi_{\sigma}^{\rm MCFM}(q)| {\cal H}^{\rm MCFM}|\Psi_{\sigma}^{\rm MCFM}(q)\rangle}{\langle \Psi_{\sigma}^{\rm MCFM}(q)|\Psi_{\sigma}^{\rm MCFM}(q)\rangle}, \nonumber \label{E_sigmaqx} \\
|\Psi_{\sigma}^{\rm MCFM}(q)\rangle&=&\prod_{\sigma}[\alpha_{0,q}\tilde{c}_{q,\sigma}^{\dagger}+\sum_{m=1}^{M}\alpha_{m,q}\tilde{d}_{m,q,\sigma}^{\dagger}]|0\rangle, \label{eq:MCFM_ground_state_q}
\end{eqnarray}
where $\beta$ is an additional variational parameter, $|\Psi_{\sigma}^{\rm MCFM}(q)\rangle$ is the eigenstate of Eq.(\ref{eq:MCFM_hamiltonian}) with the momentum $q$ and $|\Psi_{\sigma}^{\rm Hub}(q)\rangle$ is the same as Eq.(\ref{eq:Psi_Hub_qsigma}).
The product with respect to $q$ is over the full Brillouin  zone for Eq.~(\ref{eq:MCFM_var_ground_state_Fermi}).
If we take $\beta \rightarrow \infty$, it is reduced to the original algorithm using Eqs.(\ref{eq:Hub_var_ground_state2}) and (\ref{eq:Psi_Hub_qsigma}).
Therefore this Fermi machine extension certainly enhances the representability % by a nonlinear function of $|x\rangle$,
with a strategy conceptually similar to the Boltzmann machine~\cite{Carleo2017,Nomura2017} but by using a substantially different quantum algorithm.    

In the practical calculation, it is useful to represent in the matrix form:
 $|\Psi_{0 \ \sigma}^{\rm Hub}\rangle$ is represented by a $N_s \times N/2$ matrix ${\cal P}_{\sigma}$, where from Eq.(\ref{eq:Psi_Hub_qsigma}), the matrix element 
is given by
\begin{eqnarray}
({\cal P}_{\sigma})_{\ell,k}&=&   \sum_{m=0}^{M}\alpha_{m,q(k,\sigma)}%\sum_{\ell}
e^{iq(k,\sigma)\ell}(-)^{mn_{\ell,-\sigma}}.
\label{eq:Hub_var_ground_state3}
\end{eqnarray}
Here, $q(k,\sigma)$ is the momentum of the $k$th occupied electron with spin $\sigma$.
The matrix representation of the sample $|x\rangle\rightarrow {\cal X}$ is
\begin{eqnarray}
({\cal X}_{\sigma})_{\ell,\ell'}=\sum_{n}^{N_s}\sum_{i}^{N/2}\delta_{\ell,n}\delta_{\ell',i}
\end{eqnarray}
if the $i$th spin-${\sigma}$ electron is at the site $n$ in the given sample.

Since $\langle x'| {\cal H}|x\rangle$ is easily calculated in Eq.(\ref{eq:E_0ave}), here we discuss how one can calculate
$\langle \Psi_{0}^{\rm Hub}|x'\rangle$ and $\langle x|\Psi_{0}^{\rm Hub}\rangle$ in the matrix representation.
By defining $2N_s \times N$ matrices
\begin{eqnarray}
{\cal P}&=& \begin{pmatrix} {\cal P}_{\uparrow} & 0 \\ 0 & {\cal P}_{\downarrow}  \end{pmatrix}
\end{eqnarray}
and
\begin{eqnarray}
{\cal X}&=& \begin{pmatrix} {\cal X}_{\uparrow} & 0 \\ 0 & {\cal X}_{\downarrow}  \end{pmatrix},
\end{eqnarray}
we can calculate $\langle x|\Psi_{0}^{\rm Hub}\rangle$
from the determinant of the matrix product $^T{\cal X}{\cal P}$ represented by a $N\times N$ matrix, where $^T{\cal X}$ is the transpose of $\cal X$.
In Eq.(\ref{eq:Hub_var_ground_state3}), we need to be careful about the dependence on $n_{\ell,-\sigma}$.
Since $n_{\ell,-\sigma}=\sum_{k} ({\cal X}_{-\sigma})_{\ell,k}$, ${\cal P}_{\sigma}$ depends on ${\cal X}_{-\sigma}$,
but it can be easily numerically calculated, because practically, $|\Psi_{0}^{\rm Hub}\rangle$ appears only in the inner product with $\langle x|$.

In this way, $E_0$ can be calculated by Monte Carlo sampling of $|x\rangle$ as
$\sum_{x} F(x)/N_{\rm sample}$ for the important sampling generated with the probability  $w\propto |\langle x|\Psi_{0}^{\rm Hub}\rangle|^2$. Here, $N_{\rm sample}$ is the number of the samples
and 
 \begin{equation}
F(x)=\sum_{x'} \frac{{\rm det} [^T{\cal PX'} ^T{\cal X'H}{\cal X}]}{{\rm det} [^T{\cal PX}]}
\end{equation}
is estimated from the ratio of two determinants of $N\times N$ matrices. 
Then the variational parameters in Eqs.~(\ref{eq:MCFM_hamiltonian})-(\ref{eq:MCFM_Hd}) are optimized to lower $E_0$.

In the matrix representation, $|\Psi_{\sigma}^{\rm MCFM}(q)\rangle$ and $|\Psi_{\sigma}^{\rm Hub}(q)\rangle$ are $N_s$-component vectors  and for instance, $\langle \Psi_{\sigma}^{\rm MCFM}(q)|{\cal H}^{\rm MCFM}|\Psi_{\sigma}^{\rm MCFM}(q)\rangle$ is a scalar for diagonalized $N_s\times N_s$ matrix of ${\cal H}^{\rm MCFM}$ in the momentum representation of Eqs.~(\ref{eq:MCFM_hamiltonian})-(\ref{eq:MCFM_Hd}).

\begin{figure*}[tb]
		\begin{center}
			\includegraphics[width=11.4cm,clip]{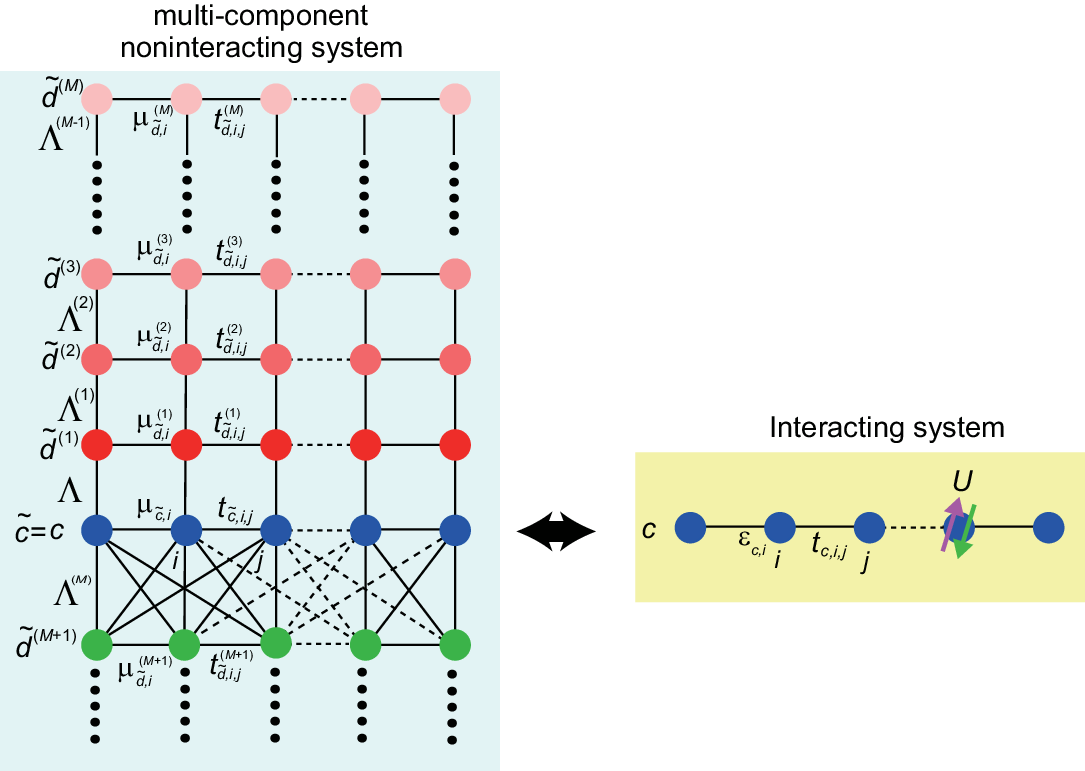}   
		\end{center}
	\caption{(Color online): Illustration of the architecture of Fermi machine network extended from Fig.~\ref{fig:mapping_MCFM} by introducing momentum dependent (spatially extended) hybridization.}
	\label{fig:extended_mapping_MCFM}
	\end{figure*}

As a further extension for more accurate variational form, long-range part of quantum entanglement may be more efficiently incorporated by adding layers connected by nonlocal hybridization such as
\begin{eqnarray}
{\cal H}_{\rm nonlocal} &=& {\cal H}^{(\tilde{d})}_{\rm nonlocal} +{\cal H}^{(\tilde{c}\tilde{d})}_{\rm nonlocal}, \label{eq:nonlocalH} \\
 {\cal H}^{(\tilde{d})}_{\rm nonlocal}  &=&\sum_{\langle i,j \rangle \sigma}(-t_{\tilde{d},i,j}^{(M+1)})(\tilde{d}_{i,\sigma}^{(M+1) \dagger}\tilde{d}_{j,\sigma}^{(M+1)}+{\rm H.c.}) \nonumber \\
&+&\sum_{i,\sigma}\mu_{\tilde{d},i}^{(M+1)}n_{\tilde{d},i,\sigma}^{(M+1)},  \nonumber \\
{\cal H}^{(\tilde{c}\tilde{d})}_{\rm nonlocal}&=&  \sum_{\sigma,\sigma'}\sum_{i, j}(\Lambda^{(M)}_{\sigma,\sigma'}(j-i)\tilde{c}_{i,\sigma}^{\dagger}\tilde{d}_{j,\sigma'}^{(M+1)} + {\rm H.c}) \nonumber \\
&=&  \sum_{\sigma,\sigma'}\sum_{k}(\Lambda^{(M)}_{\sigma,\sigma'}(k)\tilde{c}_{k,\sigma}^{\dagger}\tilde{d}_{k,\sigma'}^{(M+1)} ) \label{eq:nonlocal_hyb}
\end{eqnarray}
as is illustrated in Fig.~\ref{fig:extended_mapping_MCFM}. Here, the spin dependence is taken to satisfy the spin rotational symmetry. 
To make the correspondence between $\tilde{d}^{(M+1)}$ and $c$, one may generalize Eq.~(\ref{eq:d}) or more concretely,
$\tilde{d}_{j,\sigma'}^{(M+1)}\leftrightarrow \sum_{i,\sigma} \Theta_{\sigma',\sigma}(j-i)c_{i,\sigma} (1-2n_{j,\sigma'})$, 
which leads to the replacement of Eq.~(\ref{eq:Psi_Hub_qsigma})
with
\begin{eqnarray}
|\Psi_{\sigma}^{\rm Hub}(q)\rangle &=& 
[\sum_{m=0}^{M}\alpha_{m,q}\sum_{\ell}\frac{e^{iq\ell}}{\sqrt{N_s}}c_{\ell,\sigma}^{\dagger}(-)^{mn_{\ell,-\sigma}} \nonumber \\
&+&\alpha_{M+1,q}\sum_{\ell,\ell',\sigma'}\frac{e^{iq\ell'}}{\sqrt{N_s}}\Theta_{\sigma',\sigma}(\ell'-\ell)c_{\ell,\sigma}^{\dagger}(-)^{n_{\ell',\sigma'}}] |0\rangle,  
\nonumber \\ \label{Psi_Hub_qsigma2}
\end{eqnarray}
which means that the spin dependent $\Theta_{\sigma,\sigma'}(\ell'-\ell)$ contains the spin parallel ($\sigma=\sigma'$) and antiparallel ($\sigma=-\sigma'$) components. $\Theta_{\sigma,\sigma'}(\ell'-\ell)$ is regarded as additional variational parameters.
The present extension explicitly allows dependence of the weight of the wave function at the $\ell$th site with spin $\sigma$ on the occupation of parallel ($\sigma$) or antiparallel ($-\sigma$) component of the fermion at the $\ell'$th site, which plays a role similar to Jastrow factor in the conventional variational Monte Carlo or nonlocal coupling between the visible (physical) and hidden variables in the Boltzmann machine. However,  the present nonlocal hybridization explicitly introduces spatially extended quantum entanglement between up- and down-spin electrons.  In addition, the present scheme does not break the SU(2) symmetry in contrast to the spin Jastrow factor in the variational Monte Carlo and the Boltzmann machine.  
This is because the amplitude of $|\Psi_{\sigma}^{\rm Hub}(q)\rangle$ for the component of the up spin at the $i$th site and the down spin at the $j$th site has the same value with the opposite configuration and the matrix element of  spin diagonal component $\langle \Psi_{\sigma}^{\rm Hub}(q)|S_i^zS_j^z |\Psi_{\sigma}^{\rm Hub}(q)\rangle$ becomes the same as the off-diagonal one
$\langle \Psi_{\sigma}^{\rm Hub}(q)|S_i^+S_j^- |\Psi_{\sigma}^{\rm Hub}(q)\rangle=\langle \Psi_{\sigma}^{\rm Hub}(q)|S_i^-S_j^+ |\Psi_{\sigma}^{\rm Hub}(q)\rangle$. 
%The coefficient $\alpha_{m,q}$ is the weight of the ground state of the $M+1$ component MCFM.
This nonlocal hybridization part may also be extended to deeper layers of $M+3$rd, $M+4$th $\cdots$ up to $M+M'$th layer if necessary.

In this extension, Eqs.~(\ref{eq:d-correspondence1}) and (\ref{eq:d-correspondence2}) become more complicated, because of the nonlocality
but may enable the mutual orthogonality of $\tilde{d}_{i,\sigma}^{(m)}$ in the Hubbard terminology expected for Eq.(\ref{eq:continued_fraction_Green's_func}).
Then, if we include the $k$-independent component of $\Lambda$ in Eq.~(\ref{eq:nonlocal_hyb}), the local part represented by the first term of the r.h.s. of Eq.~(\ref{Psi_Hub_qsigma2}) can be absorbed, but still only the antiparallel spin combination is relevant for that $k$-independent component. In this sense, further generalized simpler architecture would be represented by the following unified form instead of Eqs.~(\ref{eq:Psi_Hub_qsigma}) and (\ref{Psi_Hub_qsigma2}):
\begin{eqnarray}
|\Psi_{\sigma}^{\rm Hub}(q)\rangle &=& 
\sum_{m=0}^{M}\alpha_{m,q}\sum_{\ell,\ell',\sigma'}\frac{e^{iq\ell'}}{\sqrt{N_s}}\Theta_{\sigma',\sigma}^{(m)}(\ell'-\ell)c_{\ell,\sigma}^{\dagger}
\nonumber \\
&&\times(-)^{mn_{\ell',\sigma'}} |0\rangle.
 \label{Psi_Hub_qsigma3}
\end{eqnarray}

Alternatively, one can extend in the form
\begin{eqnarray}
\frac{e^{iq\ell}}{\sqrt{N_s}}\tilde{d}^{(m)\dagger}_{\ell,\sigma}&\rightarrow & \sum_{\ell',\sigma'} \frac{e^{iq\ell'}}{\sqrt{N_s}}\Theta_{\sigma',\sigma}^{(m)}(q,\ell'-\ell)d^{(m-1)\dagger}_{\ell,\sigma} \nonumber \\
\ \ \ &&\times  (-)^{n_{\tilde{d}^{(m-1)},\ell',\sigma'}}  \\
&\rightarrow& \Xi^{(m)}_{\sigma}(q, \ell)  c_{\ell,\sigma}^{\dagger}, \\
\Xi^{(m)}_{\sigma}(q,\ell) &=& \prod_{n=1}^m \large[ \sum_{\ell_n,\sigma_n} \frac{e^{iq\ell_n}}{\sqrt{N_s}}\Theta_{\sigma_n,\sigma}^{(m-n+1)}(q,\ell_n-\ell)  \nonumber \\
&&\times (-)^{n_{\ell_n,\sigma_n}} \large], \\
\Xi^{(0)}_{\sigma}(\ell) &=&\frac{e^{iq\ell}}{\sqrt{N_s}},
\label{Psi_Hub_qsigma4}
\end{eqnarray}
which leads to
\begin{eqnarray}
|\Psi_{\sigma}^{\rm Hub}(q)\rangle &=& 
\sum_{m=0}^{M}\alpha_{m,q}\sum_{\ell}\Xi_{\sigma}^{(m)}(q,\ell)c_{\ell,\sigma}^{\dagger}|0\rangle. 
\label{Psi_Hub_qsigma5}
\end{eqnarray}

The variational optimization procedure is summarized as
\begin{itemize}
\item For given parameters in Eqs.~(\ref{eq:MCFM_hamiltonian})-(\ref{eq:MCFM_Hd}), determine $\alpha_{m,k}$ for the ground state of the MCFM and then $|\Psi_0^{\rm Hub}\rangle$ in the form of Eqs.~(\ref{eq:Psi_matrix}) and (\ref{eq:Hub_var_ground_state2}) with Eqs.~(\ref{eq:Psi_Hub_qsigma}), (\ref{Psi_Hub_qsigma2}), (\ref{Psi_Hub_qsigma3}) or (\ref{Psi_Hub_qsigma5}).
\item Calculate $ \langle x | \Psi_0^{\rm Hub} \rangle $ 
for a Monte Carlo sample given by a real space configuration $|x\rangle$ in the matrix 
representation  ${\rm det}^T{\cal X}{\cal P}$.
\item Continue the Monte Carlo sampling with the importance sampling weight  $\rho(x)=|[\langle x|\Psi_0^{\rm Hub} \rangle]|^2/\langle\Psi_0^{\rm Hub} |\Psi_0^{\rm Hub} \rangle$ proportional to $|{\rm det}[^T{\cal XP}]|^2$ and calculate the energy expectation value 
\begin{eqnarray}
E_0&=&\frac{1}{N_{\rm sample}}\sum_x F(x) \nonumber \\
&=&\frac{1}{N_{\rm sample}}\sum_{x,x'} \frac{{\rm det} [^T{\cal PX'} ^T{\cal X'H}{\cal X}]}{{\rm det} [^T{\cal PX}]}.
\end{eqnarray}
\item Optimize the variational parameters in Eqs.~(\ref{eq:MCFM_Hc})-(\ref{eq:MCFM_Hd}) to lower $E_0$ until convergence. For the optimization, the natural gradient~\cite{Amari1996,Amari1998} (stochastic reconfiguration~\cite{SorellaSR1998}) can be used.
\end{itemize}

So far, we have restricted the formulation to systems with translational invariance so that the momentum is a good quantum number. 
However, it can be easily extended to non-invariant cases such as systems under site-dependent random potential, interaction and/or bond-random hopping. In these cases, Eq.~(\ref{eq:MCFM_ground_state}) is replaced by   
\begin{eqnarray}
|\Psi_{0 \ \sigma}^{\rm MCFM}\rangle &=& \prod_k^{k_{\rm F}} |\Psi_{\sigma}^{\rm MCFM}(k)\rangle 
\label{eq:MCFM_ground_state_random}, \\
\end{eqnarray}
where $k$ is not the momentum any more and represents the $k$-th lowest eigenstate of the MCFM Hamiltonian containing randomness,
which can be obtained from the diagonalization of the MCFM Hamiltonian giving the amplitude of the eigenfunction represented by the component $\alpha_{m,\ell,k}$ for the site $\ell$ in the $m$-th layer.
Then, Eq.~(\ref{eq:Psi_MCFM_qsigma}) is replaced with
\begin{equation}
|\Psi_{\sigma}^{\rm MCFM}(k)\rangle= [\alpha_{0,\ell,k}\tilde{c}_{\ell,\sigma}^{\dagger}+\sum_{m=1}^{M}\alpha_{m,\ell,k}\tilde{d}_{m,\ell,\sigma}^{\dagger}]|0\rangle, \label{eq:Psi_MCFM_qsigma_random}
\end{equation}
and the corresponding Hubbard state (\ref{eq:Hub_var_ground_state}) is modified to
\begin{eqnarray}
|\Psi_{0 \ \sigma}^{\rm Hub}\rangle = \prod_k^{k_{\rm F}}[\sum_{m=0}^{M}\alpha_{m,\ell,k}c_{\ell,\sigma}^{\dagger}(1-2n_{\ell,-\sigma})^m]|0\rangle.
\label{eq:Hub_var_ground_state_random}
\end{eqnarray}
Further refinements discussed above may also be applied as well.
As a simple starting point one may impose a constraint of the ``quenched randomness" for the construction of the MCFM,
which means that the MCFM paramters are taken not to depend on the layer index for the same site or bond.

\section{Benchmark for 4-site Hubbard model}
\label{sec:benchmark} 
Here, the validity of the present formalism is tested in the 4-site Hubbard model at half filling and with hole doping.
The MCFM results are compared with the exact ground state given by the numerical diagonalization.
To make the parameters as simple as possible, we do not introduce the deep layers beyond the first hidden layer and the interlayer coupling $\Lambda^{(m)}$ with $m \ge 2$ are switched off in this study.  However, the nonlocal hybridization is taken into account in Sec.~\ref{sec:doped}.
The parameter $\beta$ in Eq.~(\ref{eq:f_xPsi}) is also taken to be $\infty$ so that the Fermi distribution is taken at zero temperature.
\begin{table*}[tbh]
  \caption{Comparison of estimated ground-state energies per site $E_0$ of the half-filled 4-site Hubbard model between the exact diagonalization and Fermi machine with list of examples of the optimized variational parameters.  Common (hyper)parameters are $M=1$, $M'=0$, and $t_{\tilde{c}}=1$.}
        \begin{tabular}{ l c c c c c c  }
        \hline
             & exact $E_0$  
             & $E_0$ by Fermi machine  & $t_{\tilde{d}}$ & $\Lambda$ & $\mu_{\tilde{c}}$ & $\mu_{\tilde{d}}$ \\
             \hline
            $U=4$ & -2.10275 & -2.10275 & 1.0 & -1.002 & 1.002 & 4.644\\
           $U=8$ & -1.32023 & -1.32023 & 1.0 & -2.238 & 2.238 & 7.229 \\
            & -1.32023 & -1.32023& 2.0 & -3.330 & 3.330 & 10.731 \\
\hline
        \end{tabular}
    \label{tab:4site_hf}
\end{table*}
\begin{table*}[t!]
\caption{Comparison of estimated ground-state energies per site $E_0$ of the hole doped 4-site Hubbard model between the exact diagonalization and Fermi machine with list of an example of optimized variational parameters for $U=8, t=1$ with 1 up and down doped hole each.  Parameters not listed in the table are $M=1$, $M'=1$,  $t_{\tilde{c}}=t_{\tilde{d}}=1$, $t_{\tilde{d},i,j,\sigma}^{(M+1)}=0$, and $\mu_{\tilde{d},i}^{(M+1)}=\mu_{\tilde{d},i}^{(M+1)}=8.0$.
The nonlocal hybridization is taken in a simple form as $\Lambda^{(M)}_{\sigma,\sigma'}[k]=\lambda^{(M)}_{\sigma,\sigma'}\cos [k]$ so that the interlayer hybridization is extended simply to the nearest neighbor pair of $\{i,j\}$ in Eq.~(\ref{eq:nonlocal_hyb}). Accordngly, we take $\Theta_{\sigma,\sigma'}(i-j)=1$ only for the nearest neighbor pair of ${i, j}$ for any combination of $\sigma$ and $\sigma'$ and zero otherwise
in Eq.~(\ref{Psi_Hub_qsigma2}).}  

        \begin{tabular}{ l c c c c c c c}
        \hline
             & exact $E_0$ 
             %& $\delta$ 
             & $E_0$ by Fermi machine  & $\Lambda$ & $\mu_{\tilde{c}}$ & $\mu_{\tilde{d}}$ & $\lambda^{(M)}_{\uparrow,\uparrow}=\lambda^{(M)}_{\uparrow,\uparrow}$ & $\lambda^{(M)}_{\uparrow,\downarrow}=\lambda^{(M)}_{\downarrow,\uparrow}$ \\
    \hline
            $U=8$ & -3.20775 & -3.20775 & -2.71408 & 2.71408 & 4.73336 & -0.35912 & -0.71824 \\
    \hline
    \end{tabular}
    \label{tab:4site_hd}
\end{table*}

\subsection{Half-filled 4-site Hubbard model}
The ground-state energies of the half-filled 4-site chain with the periodic boundary condition at $t=1$ are compared between the present Fermi machine using the level of Eq.~(\ref{eq:Psi_Hub_qsigma}) and the exact energies of the Hubbard model at $U=4$ and 8 in Table~\ref{tab:4site_hf}. The optimized parameters are not unique and are redundant, namely, many optimized parameters can equally reproduce the exact results. In Table~\ref{tab:4site_hf}, two examples are shown for $U=8$.
Here simple parameters similar to the 2-site case are chosen. This redundancy indicates the flexibility and representability of the present framework. 

\subsection{Doped Case} \label{sec:doped}
Table~\ref{tab:4site_hd} shows the doped case (1 up and 1 down holes doped) for $U=8$ and $t=1$. Here, the nonlocal hybridization term is necessary to reproduce the exact result. Still small numbers of parameters are enough.

\red{We note that the applicability of the present method does not have the limitation for the range of the Hamiltonian parameters including $0\le U/t \le \infty$ as is clarified in the atomic limit, the two-site system and the numerical 4-site results. }

\section{Discussion, Summary and Outlook}
In the conventional variational Monte Carlo method, the fermion Slater determinant or Pfaffian is used as the starting point~\cite{Misawa_VMCreview2019}. A crucial difference of the present Fermi machine is the entanglement with the hidden fermion degrees of freedom hybridizing with the physical (visible) fermions in the physical Hubbard model, which allows the formation of the gap structure in the spectra without spontaneous symmetry breaking, generating zeros of the Green's functions as well as poles as in the case of the genuine Mott insulator. This hybridization enables the flexible restructuring of the nodes of the wave function beyond the ``single-particle electronic structure" representable by the visible (physical) electrons for electronic systems. 
Except for the starting trial state represented by the Slater determinant or Pfaffian, the deeper level of the architecture of the conventional variational wave functions in the literature has mostly classical nature such as the Gutzwiller and Jastrow factors, which are in marked contrast with the present approach, where the deeper levels of variables have fermionic nature and entangle with the physical (visible) fermions, efficiently optimizing the node structure.  The Boltzmann machine has, in this regard, the same contrast because the hidden variable is the Ising classical spin and does not quantum mechanically entangle with the visible quantum variables. Recently proposed neural network~\cite{Moreno} introduces an unusually constrained type of fermions as the hidden variables, whose wave function is determined solely from the real-space particle configuration of the visible fermions. 
 
In the present work, the correspondence between strongly correlated fermions and multi-component noninteracting fermions is formulated. It offers an algorithm of quantum many-body solver. This Fermi machine variationaly approaches the ground state of correlated electrons by introducing dark (hidden) fermions hybridizing with the physical fermions, which substantiate the fractionalization of fermions emerging from the strong correlation effect, which has indeed been supported by the analyses of experimental results as well as the model studies of Mott insulators and high $T_c$ superconductors in the literature~\cite{Sakai2016a,Yamaji2021,Imada2021a,Singh2022,Schmid2023}.

The present correspondence establishes the relation between the ``bulk" noninteracting multi-component fermions and the ``edge"
strongly correlated fermions. This has a conceptual similarity in other correspondence in physics such as the holographic correspondence in the AdS-CFT~\cite{Maldacena,Hartnoll,Hashimoto2018}, the bulk-edge correspondence in topological materials~\cite{Hasan,Ando} and the mapping of $d$-dimentional quantum systems to $d+1$-dimensional classical ones in the path integral, though physical contents are substantially different. Further  deep connections to these three frontiers are interesting future issues.

The efficiency of the present formalism for larger system sizes for a practical use is a future important subject to be pursued.
For this purpose, efficient optimization of the variational parameters must be examined and tested, which is left for future studies.

In this paper, we have examined the ground state of the noninteracting MCFM represented by a Slater determinant. However, in the conventional variational Monte Carlo, pair-product wave functions instead of Slater determinants show higher accuracy~\cite{Misawa_VMCreview2019}.  It is an intriguing future issue, whether the mapping from ground states of the Hartree-Fock-Bogoiubov Hamiltonian, namely, a mean-field ground state of the symmetry broken Hamiltonian such as superconducting or magnetic mean-field order shows higher accuracy. 

%\if0
{\bf Acknowledgements}
The author is grateful to Wei-Lin Tu for useful comments. The author also thanks discussions with Filippo Vicentini, Ryui Kaneko and Shiro Sakai. This work is financially supported by MEXT KAKENHI, Grant-in-Aid for Transformative Research Area (GrantNo. JP22H05111 and No.JP22H05114).
This work is also supported by MEXT as ``Program for Promoting Researches on the Super computer Fugaku" (Simulation for basic science: approaching the new quantum era, Grant No. JPMXP1020230411).
%\fi

\appendix
\setcounter{secnumdepth}{1}
\section{Repsentability of the ground state} \label{appendix:Representability}

In this Appendix we discuss the representability of the wave function by the present scheme.
When we exchange the order of the product with respect to $q$ and the summation over $m$ in Eqs.~(\ref{eq:Hub_var_ground_state2}) and (\ref{eq:Psi_Hub_qsigma}), we obtain
\begin{eqnarray}
|\Psi_{0 \ \sigma}^{\rm Hub}\rangle &=& 
\sum_{m=0}^{M} |\Psi_{\sigma,m}^{\rm Hub}\rangle,
\label{eq:Hub_var_ground_state2_rep} \\
|\Psi_{\sigma,m}^{\rm Hub}\rangle &=&
\prod_q^{k_{\rm F}}[\alpha_{m,q}\sum_{\ell}\frac{e^{iq\ell}}{\sqrt{N_s}}c_{\ell,\sigma}^{\dagger}(-)^{mn_{\ell,-\sigma}}]|0\rangle. \nonumber \\
\label{eq:Psi_Hub_qsigma_rep}
\end{eqnarray} 
This can be interpreted as the representation by a linear combination of $M$ Slater determinants, where the $m$-th one is $|\Psi_{\sigma,m}^{\rm Hub}\rangle$, aside from the entanglement of $\sigma$ and $-\sigma$ electrons contained in the odd $m$ terms. This exchange becomes justified in the following setup: By taking $\alpha_{m,q}=0$ for the odd $m$, which is the case of $\mu_{\tilde{d},i}^{(m)}=\infty$ for odd $m$, the even $m$ layers are indeed decoupled each other and each multi-particle state becomes a simple Slater determinant consisting of the single-layer component with even $m$.  Then if the Hamiltonian parameters of each even $m$th layer is taken in such a way that the ground-state energy of each single-layer Slater determinant is degenerate with those of different even $m$th layers, we may take any linear combination of them, generating a multi-Slater determinant. (Note that in this case, the weights of the linear combination are the variational parameters as well.)  
The construction of the MCFM Hamiltonian parameters to satisfy the degeneracy is of course possible, because the MCFM energy and occupied momenta at each layer can be freely tuned by taking appropriate dispersion at each layer.  Since the complete set of the Hilbert space can be expanded by linear combinations of fermion determinants, %(namely, by multi-Slater determinant), 
this is a formal proof of the representability %of any eigenfunction 
of the Hubbard model (more generally, any interacting lattice fermion system). 

Of course, the number  of Slater determinants for the complete representability increases rapidly with increasing system size if one follows this naive muli-determinant representation and this proof is just formal, while practically, the increase can be suppressed by a proper choice of the odd $m$ contribution as well as by the spatially off-diagonal hybridization, as is discussed from Eq.(\ref{eq:nonlocalH}) through Eq.~(\ref{Psi_Hub_qsigma5}).  

More importantly, Eq.~(\ref{eq:Psi_Hub_qsigma}) %and its extension 
allows the sign change of the wave function dynamically when the opposite spin is occupied. This flexibly adjusts the nodal structure of the fermion wave function required for interacting systems. 

\bibliographystyle{jpsj_mod}
%\bibliography{Ref_Imada}

\end{document}